# Geometric tuning of charge and spin correlations in manganite superlattices

K. Rogdakis[1,a),b)], Z. Viskadourakis[1,2], A.P. Petrović[3], E. Choi[4], J. Lee[4] and C. Panagopoulos[1,2,3,5,b)]

[1] Institute of Electronic Structure & Laser, Foundation for Research and Technology-Hellas, Vassilika Vouton, Heraklion 71110, Greece

[2] Cretan Center for Quantum Complexity and Nanotechnology, University of Crete, Heraklion 71003, Greece

[3] School of Physical and Mathematical Sciences, Division of Physics and Applied Physics, Nanyang Technological University, 637371 Singapore

[4] School of Advanced Materials Science and Engineering, Sungkyunkwan University, Suwon 440-746, Republic of Korea

[5] Department of Physics, University of Crete, Heraklion 71003, Greece

[a)] Current address: CNRS, Institut Néel, B.P. 166, 38042 Grenoble Cedex 09, France
[b)] Authors to whom correspondence should be addressed. Electronic mails: rogdakis@iesl.forth.gr and christos@ntu.edu.sg

**Abstract**

We report a modulation of the in-plane magnetotransport in artificial manganite superlattice (SL) $[(NdMnO_3)_n /(SrMnO_3)_n /(LaMnO_3)_n]_m$ by varying the layer thickness $n$ while keeping the total thickness of the structure constant. Charge transport in these heterostructures is confined to the interfaces and occurs via variable range hopping (VRH). Upon increasing $n$, the interfacial separation rises, leading to a suppression of the electrostatic screening between carriers of neighboring interfaces and the opening of a Coulomb gap at the Fermi level ($E_F$). The high-field magnetoresistance (MR) is universally negative due to progressive spin alignment. However at a critical thickness of $n$=5 unit cells (u.c.), an exchange field coupling between ferromagnetically ordered interfaces results in positive MR at low magnetic field ($H$). Our results demonstrate the ability to geometrically tune the electrical transport between regimes dominated by either charge or spin correlations.

Transition metal oxides are of technological interest due to their unique variety of functional properties which are tunable by a wide range of parameters.[1,2] Selective chemical doping,[3] structural changes,[4] controllable strain[5] and the application of external electromagnetic fields[6] are just a few of the myriad methods which enable materials to be switched between insulating, semiconducting or metallic phases, often revealing emergent phenomena such as ferroelectricity,[7] magnetism,[8,9] colossal magnetoresistance[4,10,11] or high temperature superconductivity[12]. Although this extreme sensitivity to external parameters initially posed difficulties due to inhomogeneity during material synthesis, recent advances in oxide thin film deposition have resulted in a structural quality rivaling that of the best semiconductors.[1,2,13] In particular, the ability to prepare extended series of atomically abrupt and structurally coherent interfaces between epitaxial oxide films has greatly facilitated the simultaneous tuning of charge, spin, orbital and lattice degrees of freedom, often leading to novel behavior absent from the bulk components of the structure.[1,7,9,13,14]

The artificial SL [$(NdMnO_3)_n$ / $(SrMnO_3)_n$ / $(LaMnO_3)_n$ ]$_m$, consisted of three different manganites (tri-layer SL), provides a textbook example of emergent behavior. It is comprised of insulating antiferromagnetic (AFM) manganite[15] layers in a ($n,m$) configuration, where index $n$ defines the layer thickness (in u.c.) of each manganite, while $m$ is the total number of tri-layers in the SL. $SrMnO_3$ (a G-type antiferromagnet with $Mn^{3+}$ valency) is sandwiched between $NdMnO_3$ and $LaMnO_3$ layers (which are both A-type antiferromagnets with $Mn^{4+}$ valency).[15] Charge transfer is expected at the $NdMnO_3$/$SrMnO_3$ and $SrMnO_3$/$LaMnO_3$ interfaces due to polar discontinuity, leading to a mixed interfacial valency.[16] Previously, we have shown that although the three constituent manganites are non-ferroic, the asymmetric SLs display ferroelectric behavior perpendicular to the interfaces which may be tuned by an external $H$ as well as by varying $n$.[7] We have also observed magnetic relaxor and spin glass-like properties at the interfaces, which exhibit a similar dependence on $n$.[9] Here, we advance these studies by investigating the evolution of the in-plane electrical transport with the geometric control parameter $n$. We find that the MR magnitude and sign, as well as the conduction type vary upon tuning the interfacial separation which is the result of the interplay between spin and charge correlations. Geometrical tuning of the SL therefore provides an attractive handle to switch between distinct spin and charge correlation-dominated transport regimes.

Laser molecular-beam epitaxy is used to grow $[(NdMnO_3)_n / (SrMnO_3)_n / (LaMnO_3)_n]_m$ SLs on single-crystalline $SrTiO_3$.[17] Details regarding the material growth, and structural, magnetic and electrical characterizations can be found in the supplementary material[18] as well as in our previous studies.[7,9,17] Three different SL configurations were used, namely $(n,m)$ = *(2,21)*, *(5,8)* and *(22,2)*, thus maintaining a constant total SL thickness of approximately 50 nm. A 3D schematic of a typical sample is shown in the inset of Fig. 1.

Figure 1 shows the temperature-dependent in-plane resistance $R_s(T)$ for all three SLs between 60K and 200K.[18] Previous studies have determined that the physical properties of these SLs are dominated by the interfaces:[7,9] at 200K $R_s$ is roughly proportional to the number of interfaces in each SL, thus supporting this assertion. $R_s$ increases with decreasing temperature, indicating the insulating nature of all three SLs. Certain $LaMnO_3 / SrMnO_3$ bi-layer SLs have been reported to become metallic for layer thicknesses $n < 5$.[19,20] However, optimized film growth was shown to result in insulating behavior even for very thin layers, by suppressing the interfacial rugosity and defects responsible for extrinsic metallic conductivity.[21] The insulating behavior of our samples – especially the *(2, 21)* SL – is therefore testament to their high structural quality. Upon applying an in-plane $H$ ($H_{\parallel}$) equal to 6T, a negative MR gradually develops in all the SLs below 200K. Negative MR is a common feature of mixed valence manganites and indicates a predisposition towards ferromagnetic (FM) ordering.[15]

Transport in mixed-valence manganites is known to be influenced by magnetic disorder[15], while atomic-scale defects at epitaxial interfaces are also expected to generate charge disorder by localizing electrons. At high temperatures, transport occurs via dielectric polaron hopping, crossing over to VRH for decreasing temperature or rising disorder.[15,22] We therefore analyse our $R_s(T)$ within the framework of Mott VRH theory[23]: $R = R_0 \exp((T_0/T)^\nu)$, where $\nu = (1+d)^{-1}$, $1 \leq d \leq 3$ is the effective dimensionality (e.g. $\nu=0.25$ in a 3D material), while $T_0$ is the characteristic Mott temperature which is proportional to the VRH hopping energy. In the presence of strong electron-electron interactions (i.e. a large Coulomb repulsion), a quadratic gap in the density of states opens at the $E_F$ and we enter the Efros-Shklovskii (ES) VRH regime in which $\nu=0.5$ regardless of the dimensionality.[24,25] For real materials, $\nu$ is known to deviate from the standard Mott or ES-VRH values[26], particularly in $H$[27]. We therefore set $\nu$, $R_0$, $T_0$ as free parameters and perform least-squares fits to our $R_s(T)$ data in the 60K – 150K temperature range;[18] the results are shown in Figs. 2(a)-2(f). An attempt to model our data using a dielectric polaron hopping model[28] was unsuccessful (Fig. 2(c)): we attribute this failure to the essential role of disorder in the interfacial transport, combined with the suppression of polaron formation at low temperature.[22]

The evolution of the VRH exponent $\nu$ with interfacial spacing and $H$ is summarized in Fig. 2(g). We first examine the zero-field results: for the *(22,2)* SL, $\nu(0T) = 0.52$, close to the $\nu = 0.5$ of ES VRH. In contrast, the *(2,21)* SL exhibits a markedly lower $\nu(0T) = 0.43$, although this value still lies above the 2D VRH exponent ($\nu = 0.33$). The reduction of the VRH exponent $\nu$ is a direct consequence of interfacial electronic transport.[20,22] For large $n$ (widely-spaced interfaces), the in-plane electron-electron interactions are dominated by Coulomb repulsion, leading to ES VRH conduction. Conversely, for small $n$ the interfaces are closely spaced and hence the effective 3D carrier density rises, which screens the Coulomb repulsion and partially suppresses the gap at $E_F$. The VRH exponent therefore falls towards its 2D value.[29] Varying $n$ therefore provides us with a handle to tune the effective Coulomb interaction in our SLs and we naively expect $0.43 < \nu(0T) < 0.52$ for the *(5,8)* SL. However, *(5,8)* SL does not follow this trend, exhibiting $\nu(0T) = 0.64$. This zero-field hopping exponent is unusually large and approaches the $\nu = 0.67$ predicted for ES-VRH transport in a magnetic field.[27] This strongly suggests the presence of an internal field (i.e. long range magnetic order) at the interfaces of the (*5,8*) SL.

In a $H_{\parallel}=6T$, the VRH exponent in the *(22,2)* SL rises to 0.73, close to the expected value for ES VRH in $H$ ($\nu = 0.67$).[27] A small rise in $\nu$ is also observed for the *(2,21)* SL, in qualitative agreement with the expected jump from 0.33 to 0.5 for 2D Mott VRH in $H$.[26,27,30] However, $\nu(6T)$ in the *(5,8)* SL is reduced from its zero-field value. The apparent dichotomy between the *((2,21),(22,2))* and the *(5,8)* SLs is further demonstrated by considering the evolution of $T_0(H)$ ((Fig. 2(h)): while $T_0$ falls at $H_{\parallel}=6T$ for the *(22,2)* and *(2,21)* SLs, the *(5,8)* SL again defies the trend by exhibiting a small rise in $T_0(H)$. To understand this behavior, we consider the $H$-dependence of the hopping energy in magnetic and non-magnetic disordered materials. In a non-magnetic disordered semiconductor, $H$ causes real-space shrinkage of the electron wavefunctions transverse to $H$ orientation.[24] The electrons are therefore increasingly localized, leading to a rise in the hopping energy and hence $T_0 \alpha H$. This is a purely

Coulombic effect, i.e., it is unrelated to spin correlations. Conversely, in a disordered magnetic semiconductor the spin correlations become relevant to the transport. Above the Curie temperature $T_c$, $T_0$ falls as $H$ rises (since $T_0(H) \propto T_0(1-(M(H)/M_s)^2)$, where $M(H)$ is the magnetization and $M_s$ the saturation magnetization).[15] Below $T_c$, the Coulomb interaction will once again dominate the transport and hence $T_0 \propto H$. Our previous magnetization measurements presented in ref. 9 and the high zero-field VRH exponent $\nu = 0.64$ both indicate that the *(5,8)* SL is magnetically ordered, and we observe $T_0$ to rise with $H$ as expected. In contrast, the *(2,21)* and *(22,2)* SLs exhibit paramagnetic behavior but no long-range order: consequently, $T_0$ falls as $H$ increases due to progressive spin alignment with the applied $H$. Magnetotransport in our SLs is therefore determined by the competition between wavefunction shrinkage (a purely Coulombic effect) and the magnetization of the interface (which is governed by spin correlations). Notably, geometric control of the magneto-transport properties was previously reported for a Ca-based manganite SL due to structural changes resulting from phase separation.[4]

Supporting information concerning this delicate interplay between spin and charge correlations may be obtained from isothermal MR measurements. In Figs. 3(a)-3(c), we plot in-plane MR curves at a range of temperatures for all three SLs, summarizing the temperature dependence of the MR in Figs. 3(e) and 3(f) at $H_\parallel$=0.5T and 6T, respectively. The *(22,2)* and *(2,21)* SLs (Figs. 3(a),3(c)) display qualitatively similar features: a strongly negative MR which varies monotonically with $H_\parallel$ and whose low-field gradient steepens as the temperature falls. At low temperatures and large $H_\parallel$, the MR begins to saturate; this is particularly clear in the *(2,21)* SL which exhibits the largest negative MR (>90% at 6T).

The MR of the *(5,8)* SL is shown in Fig. 3(b). Although negative, its amplitude is markedly lower than those of the other two SLs. At low temperatures (T < 50K), the negative MR varies almost linearly with *H*, without any tendency towards saturation: such behavior is characteristic of progressive spin alignment in canted or frustrated FM domains due to competing AFM/FM interactions.[24] Upon increasing the temperature, symmetric positive cusps emerge in the MR at $H_\parallel = \pm 0.5$T for 70K < *T* < 130K. This temperature range closely corresponds to the Néel temperatures of $NdMnO_3$ (80K) and $LaMnO_3$ (130K). The appearance of this cusp is highlighted in Figs. 3(d),3(e): while the negative MR in the *(2,21)* and *(22,2)* SLs increases monotonically with decreasing temperature, a local maximum corresponding to a positive MR regime appears in the 70 - 130K range for the *(5,8)* SL. We examine the anisotropy of the MR cusp shown by the *(5,8)* SL in Figs. 4(a)-4(d). The perpendicular MR $R(H_\perp)$ also shows a cusp in a similar temperature range to $R(H_\parallel)$; however it is centred at higher fields $H_\perp \approx 2$T and its amplitude peaks at lower temperature (~80K vs. ~90K). This indicates that the interfaces in the *(5,8)* SL are magnetically anisotropic, with the spins preferentially lying in-plane.

A similar positive MR cusp at low *H* was previously reported in $LaMnO_3$/$SrMnO_3$ bi-layer SLs.[19,20] The cusp was visible at temperatures as low as 10K, but only appeared for specific layer thickness: $(LaMnO_3)_{10}$ / $(SrMnO_3)_5$ and $(LaMnO_3)_3$ / $(SrMnO_3)_6$. The common feature of these heterostructures and our *(5,8)* SL, is the thickness of the $SrMnO_3$ layer: 5-6 u.c.. However, in our *(5,8)* SL the cusp is only visible in the 70 - 130 K temperature range. The AFM transition of the $NdMnO_3$ layer therefore seems to "switch off" the mechanism responsible for the cusp formation. Similar MR cusps have been also found in other manganite SLs and attributed to an AFM exchange coupling between FM regions.[31-33] In our *(5,8)* SL, a coupling of this form implies the presence of an exchange field at the interface: to quantify this field we follow the method of Krivorotov et al.[32] and express the total $H$, $H_T$, as the sum of the applied ($H_A$) and exchange ($H_{EX}$) fields: $H_T = (H_{EX}^2 + H_A^2 - 2 H_{EX} H_A m(H_A))^{1/2}$, where $m(H_A)$ is the SL magnetization normalized to its saturation magnetization. Determining $m(H_A)$ experimentally, we model our MR data using a power-law $R(H_T) = aH_T^b+c$, where $a$, $b$, $c$ and $H_{EX}$ are free parameters (Figs. 4(e),4(f)). This simple exchange field model provides an accurate reproduction of our data with $H_{EX} = 0.69$T and 0.54T for 90K and 120K, respectively. Since the eventual presence of the MR cusp is acutely sensitive to the SrMnO$_3$ layer thickness, we suggest that the interfacial exchange field is mediated via direct exchange through the SrMnO$_3$. Interestingly, in ref. 33 the authors influenced the magnitude of the coupling by using interlayers of SrTiO$_3$ and BaTiO$_3$ between La$_{0.7}$Sr$_{0.3}$MnO$_3$ and SrRuO$_3$. Perhaps the ferroic orders in BaTiO$_3$ and NdMnO$_3$, which serves as an interlayer between (LaMnO$_3$) / (SrMnO$_3$) in our SLs, play an active role in the suppression of the exchange coupling observed in the two systems but in a distinctive way. However, further work to elucidate the temperature-dependent microscopic spin structure of these SLs will be necessary for a full understanding of their magnetotransport behavior.

The experimental evidence presented above demonstrates that the delicate interplay between spin and charge correlations may be tuned by $n$. Figure 5 summarizes the evolution of the physical mechanisms controlling the transport. For large $n$, the interfaces are sufficiently far from each other to be regarded as non-interacting (Fig. 5(c)). Charge transport is therefore dominated by the in-plane Coulomb repulsion, leading to the observation of ES VRH behavior in the *(22,2)* SL. In contrast, when the interfaces are closely spaced ((Fig. 5(a)), the Coulomb interaction is partially screened due to charge correlations between neighboring interfaces. Experimentally, this results in a reduction of the VRH exponent towards the value expected for localized carriers in 2D. A large yet tunable negative MR indicates a tendency towards FM ordering in both the low and high interfacial spacing limits, but there is little magnetic interaction between interfaces. At an intermediate layer spacing $n$=5 (Fig. 5(b)), a strong exchange interaction develops between ferromagnetically-ordered interfaces, corresponding to an exchange field of at least 0.69T at 90K. This drives a crossover from negative to positive MR at low $H$. We anticipate that this ability to precisely tailor the electronic properties of atomic-scale conducting channels will prove invaluable in the quest to develop functional oxide heterostructures.

We thank J. Seo, J. D. Burton and E. Y. Tsymbal for insightful discussions. This work was supported by the National Research Foundation, Singapore through a Competitive Research Program grant NRF-CRP4-2008-04 and the European Union through EURYI, MEXT-CT-2006-039047 and FP7-REGPOT-2012-2013-1 (grant agreement No 316165) grants. The work at Sungkyunkwan University was supported by the National Research Foundation of Korea through Basic Science Research Program (2009-0092809).

**Figure Captions**

FIG. 1. In-plane $R_s(T)$ for all studied SLs. Solid lines correspond to data acquired for $H_\| = 0$, while open circles display $R_s(T)$ in $H_\| = 6$ T. For clarity, the *(5, 8)* SL data have been multiplied by 5. A schematic view of a typical Van der Pauw device used in this investigation is shown in the inset.

FIG. 2. (a)-(f) $\ln(R_s)$ vs. $T^{-\nu}$ (broad colored lines) for each SL, acquired in H = 0 (left column) and $H_\| = 6$T (right column). Solid lines correspond to VRH fits. For the *(2, 21)* SL in H=0 (c), a least-squares fit to a polaron hopping model $R_s \alpha\ T\exp(b/k_BT)$ is also shown in red. (g) Evolution of the VRH exponent $\nu=(1+d)^{-1}$ from H=0 to H=6T, shown for all SLs. (h) Characteristic VRH temperature $T_0$ at H=0 and H=6T for all SLs. Dotted lines in (g) and (h) serve as guides for the eye.

FIG. 3. In-plane MR at various temperatures for each SL: (a) *(22, 2)*, (b) *(5, 8)* and (c) *(2,21)*. (d) Low field in-plane MR for the *(5, 8)* SL at selected temperatures. Temperature dependence of the MR in each SL for $H_\| = 0.5$ T (e) and 6 T (f).

FIG. 4. (a)-(d) Comparison between the perpendicular and parallel MR in the *(5, 8)* SL at various temperatures. In-plane magnetization and MR of *(5, 8)* SL measured at 90K (e) and 120K (f).

FIG. 5. Cartoon schematic illustrating the evolution in physical properties of manganite SLs upon geometric tuning. The interfaces at which charge transport occurs are highlighted in transparent red shading.

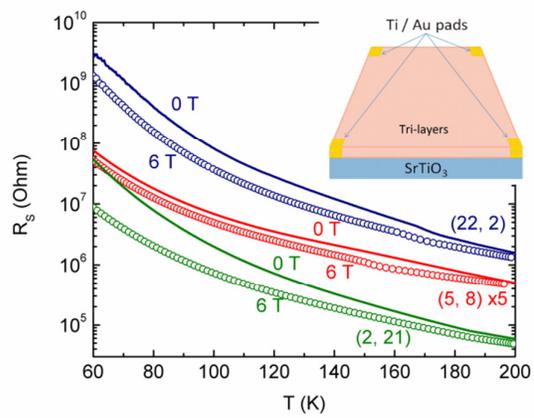

FIG. 1

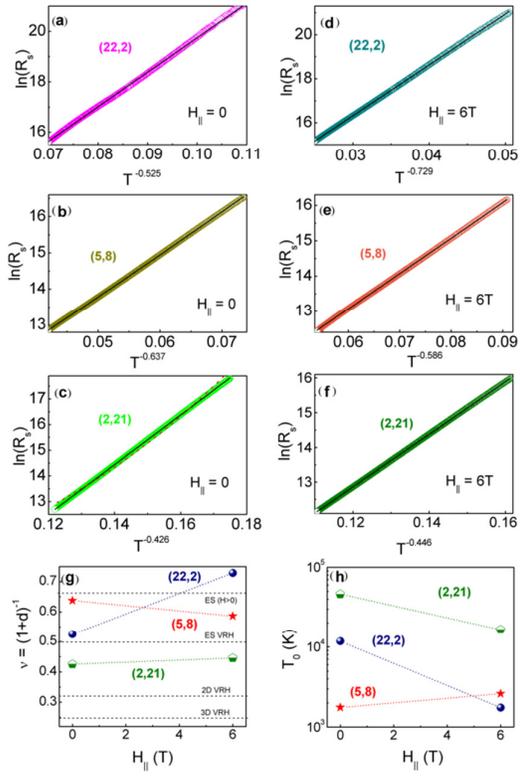

FIG. 2

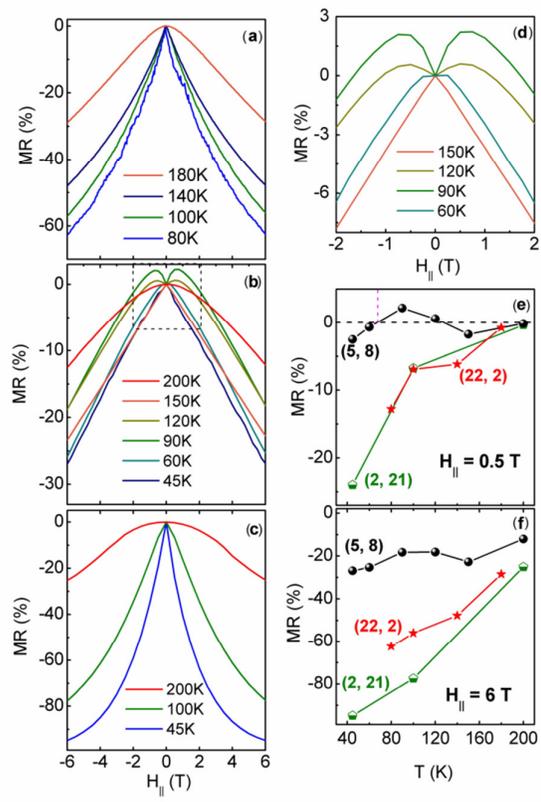

FIG. 3

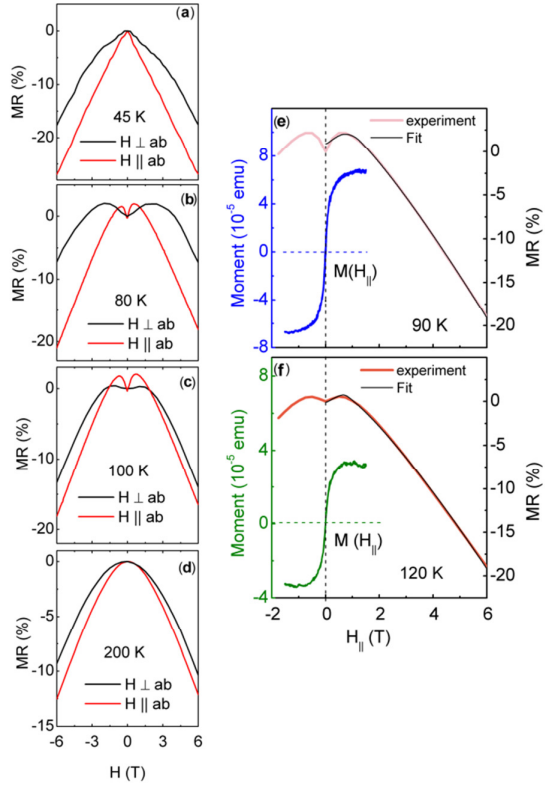

FIG. 4

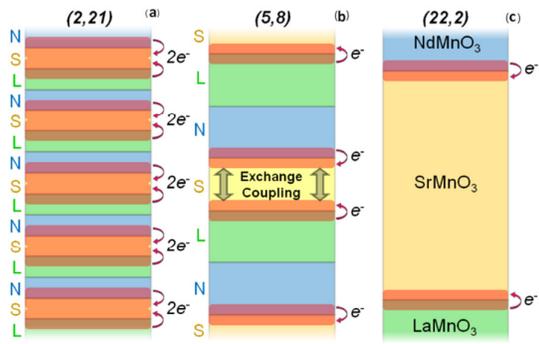

FIG. 5